\documentclass{cs20proc}

\usepackage{amsmath, amssymb}
\usepackage{bm, mathtools, cancel, empheq, mathrsfs, natbib}
\usepackage{graphicx}

\newcommand{\curl}{\nabla\times}
\newcommand{\rsun}{R_\odot}

\newcommand{\rbcz}{r_{\rm{BCZ}}}

\newcommand{\cp}{c_{\rm{p}}}
\newcommand{\pderiv}[2]{\frac{\partial#1}{\partial#2}}
\newcommand{\sn}[2]{#1\times10^{#2}}

\newcommand{\rhobar}{\overline{\rho}}
\newcommand{\av}[1]{\langle#1\rangle}

\newcommand{\e}{\hat{\bm{e}}}
\newcommand{\prm}{\rm{Pr_m}}

\newcommand{\domcz}{\Delta\Omega_{\rm{CZ}}}
\newcommand{\domrz}{\Delta\Omega_{\rm{RZ}}}

\newcommand{\prot}{P_{\rm{rot}}}
\newcommand{\bpol}{\bm{B}_{\rm{pol}}}
\newcommand{\tes}{t_{\rm{ES}}}

\editors{S. J. Wolk, A. K. Dupree, H. M. G\"unther}
\publisher{Zenodo}
\conference{The 20.5th Cambridge Workshop on Cool Stars, Stellar Systems, and the Sun}
\conferencedate{2021}

\title{Building and Maintaining a Solar Tachocline through Convective Dynamo Action}
\author{Loren I. Matilsky$^{1}$,
        Juri Toomre$^{1}$}

\affiliation{$^{1}$ JILA \& Department of Astrophysical and Planetary Sciences, University of Colorado, Boulder, CO, USA}

\shorttitle{Solar Tachocline from Convective Dynamo}
\shortauthors{Loren I. Matilsky \& Juri Toomre}

\abs{For more than thirty years, the dynamical maintenance of the thin solar tachocline has remained one of the central outstanding problems of stellar astrophysics. Three main theories have been developed to explain the tachocline's thinness, but so far none of them has been shown to work convincingly in the extreme parameter regime of the solar interior. Here, we present a rotating, 3D, spherical-shell simulation of a combined solar-like convection zone and radiative zone that achieves a tachocline built and maintained by convective dynamo action. Because of numerical constraints, the dynamo prevents the viscous spread of the tachocline instead of the Eddington-Sweet-time-scale radiative spread believed to occur in the Sun. Nonetheless, our simulation supports the scenario of tachocline confinement via the cyclic solar dynamo, and is the first time one of the main confinement scenarios has been realized in a global, 3D, spherical-shell geometry including nonlinear fluid motions and a self-consistently generated dynamo.}

\begin{document}

\maketitle

\section{Introduction}
\begin{figure}
	\centering
	\includegraphics[width=0.85\linewidth]{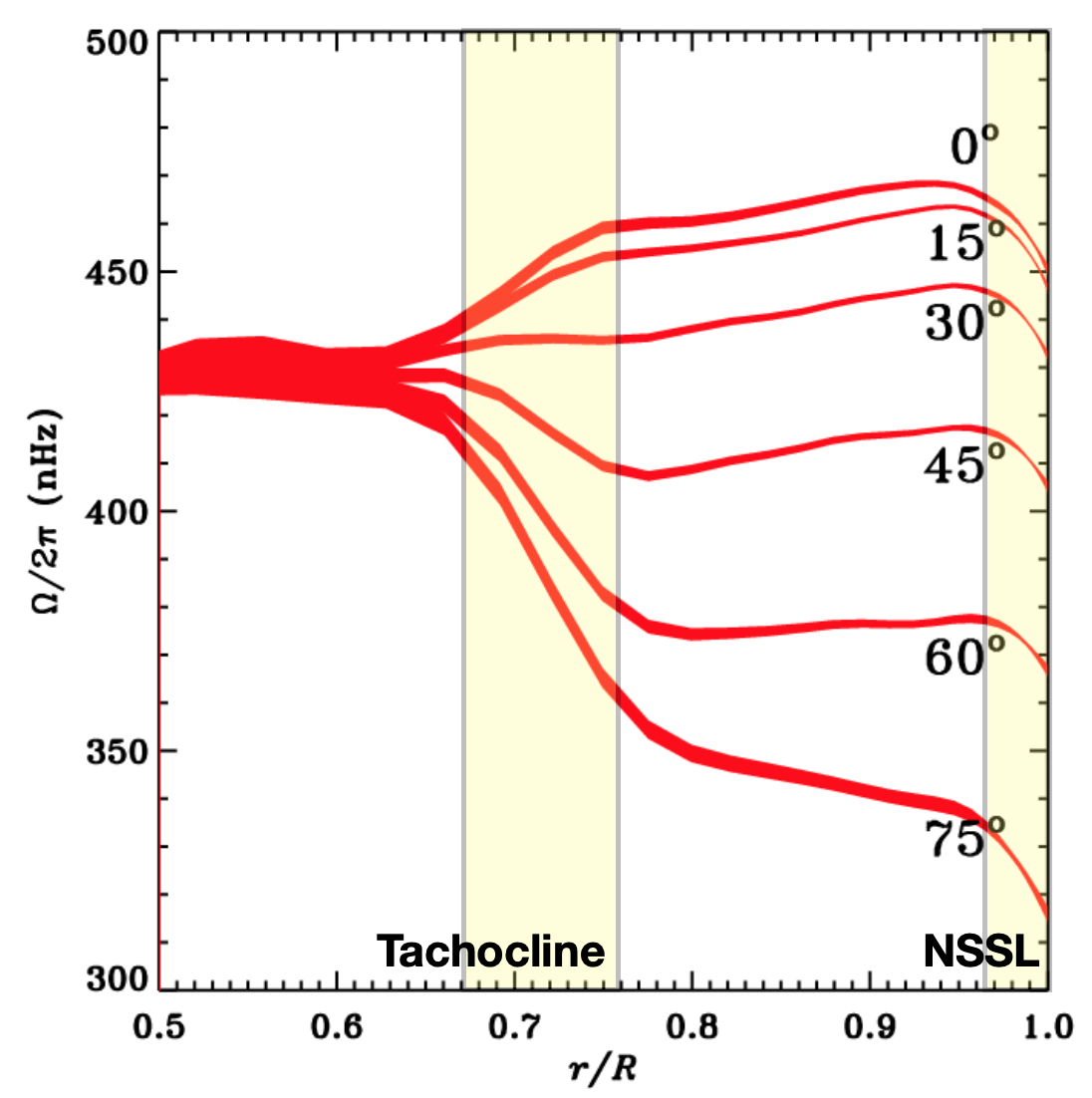}
	\caption{The internal solar rotation rate as measured by helioseismology, averaged in longitude and time (1995--2009). Rotation rate is plotted as a function of fractional solar radius at different latitudes. The width of a given curve represents the measurement error. The two shear layers at the top and bottom of the convection zone (CZ) are highlighted. Image credit: \citet{Howe2009}.}
	\label{fig:solar}
\end{figure}

The internal solar rotation rate revealed by helioseismic probing is shown in Figure \ref{fig:solar}. Helioseismology can detect the rotation rate with good precision in the whole convection zone (CZ; $r\gtrsim\rbcz\equiv0.72\rsun$, where $r$ is the local solar radius and $R_\odot\approx\sn{6.96}{10}\ \rm{cm}$ is the full solar radius) and part of the radiation zone (RZ) below. The most striking features are the two shear layers at the top and bottom of the CZ. In the \textit{near-surface shear layer} (NSSL; $r\gtrsim0.95\rsun$), the rotation rate falls sharply (by $\sim$5\%) with increasing radius. The dynamical origins of the NSSL are still not well-understood (for recent work, see \citealt{Hotta2015, Matilsky2018, Matilsky2019, Choudhuri2021}) and it may play a major role in the solar dynamo \citep{Brandenburg2005a}. 

At the base of the CZ, the \textit{solar tachocline} is also a region of intense shear, separating the differential rotation of the CZ from the nearly solid-body rotation of the RZ (for recent reviews, see \citealt{Miesch2009}, \citealt{Brun2019}, and references therein). The tachocline's proximity to the RZ makes it a keystone element of mean-field dynamo theories. The tachocline likely smooths poloidal magnetic field into large-scale ``wreaths'' of toroidal magnetic field (e.g., \citealt{Browning2006, Hathaway2015}), which can then be stored in the largely motion-free RZ. In this picture, the RZ acts as a ``magnetic reservoir,'' housing large-scale toroidal flux ropes until they are intense enough to buoyantly erupt to the surface, forming sunspot pairs (e.g., \citealt{Miesch2005}). Furthermore, the stability of these wreaths with respect to rotation and buoyancy (magnetorotational instability) may explain the emergence of sunspots primarily at latitudes $\lesssim30^\circ$ \citep{Gilman2017, Gilman2018}. It is thus of central importance to accurately assess how the tachocline is dynamically maintained, and how it interacts with the solar magnetic field and intensely turbulent convection. 

At face value, the heliosiesmic results give a tachocline thickness of $\Delta\sim$ 0.1 $\rsun$, as shown in Figure \ref{fig:solar}. This largely represents the limited spatial resolution of the helioseismic inversion kernels, which broadens with increasing depth. In reality, the tachocline may be quite thin ($\Delta\lesssim0.02\rsun$; see \citealt{Elliott1999}). The thin solar tachocline was a surprise for dynamicists. In \citet{Spiegel1992}, it was shown that the thermal wind (hot pole, cool equator) associated with the CZ's differential rotation efficiently imprints onto the RZ via radiative diffusion. The thermal wind carries with it a large-scale meridional circulation that transports angular momentum and eventually forces the RZ to rotate differentially. This so-called \textit{radiative spreading} phenomenon should have imprinted the differential rotation down to at least $r\sim0.3\rsun$ by the current age of the Sun. 

Accordingly, three major theories were developed to explain the counteraction of radiative spreading to keep the solar tachocline confined to a thin layer. The \textit{fast} theory \citep{Spiegel1992} supposes that fluid instabilities in the stably statified RZ lead to an anistropic turbulent viscosity, which is much stronger in the horizontal directions than the radial direction. This viscosity will tend to efficiently eliminate any latitudinal differential rotation imprinted from above on the dynamical timescale of $\sim$30 days (i.e., one solar rotation period). For a recent review of the fast confinement scenario, see \citet{Garaud2020} and references therein. The \textit{slow magnetic} theory \citep{Gough1998} invokes a weak poloidal magnetic field remnant in the RZ. As the inward-burrowing meridional circulation imprints differential rotation downward, the associated shear generates a back-reacting magnetic torque to prevent further spread. This process operates on the Eddington-Sweet timescale of $\tes\sim\sn{2.2}{11}$ years (see \citealt{Wood2018} and references therein for a review). Finally, the \textit{fast magnetic} theory \citep{ForgcsDajka2001} follows a similar argument to \citet{Gough1998}, but the necessary poloidal field, instead of being a primordial remnant, comes from the cyclic magnetic field associated with the solar dynamo penetrating to a skin-depth below the base of the CZ. The fast magnetic scenario operates on the period of the solar polarity cycle, namely $\sim$22 years. 

Here, we discuss a rotating, spherical-shell dynamo simulation that magnetically builds and maintains a solid-body-rotating RZ and differentially-rotating CZ, with a tachocline-like shear layer in between. As a whole, our simulation is similar to the fast magnetic confinement scenario, and is the first time any of the main scenarios have been successfully achieved in a global, 3D simulation with a self-consistent dynamo. However, we point out that numerical constraints prohibit us from running our simulation on the very long Eddington-Sweet timescale (or in the highly turbulent regime appropriate to a stellar interior), and so we cannot say with certainty whether the fast magnetic scenario is indeed taking place in the Sun. Nonetheless, simulations like the one presented here provide convincing evidence that the fast magnetic confinement scenario is possible outside a 1D or 2D model without convection (like the models explored in, e.g., \citealt{ForgcsDajka2002, Barnabe2017}).

\section{Dynamo Simulation of Solar CZ and RZ}\label{sec:exp}

We consider the anelastic (magneto)hydrodynamic---(M)HD---equations in a global, 3D, rotating spherical shell representing most of the solar CZ and a portion of the underlying RZ (in particular, the radial interval between $0.49\rsun$ and $0.95\rsun$). We solve these equation numerically using the open-source, pseudo-spectral code Rayleigh 0.9.1 \citep{Featherstone2016a, Matsui2016, Featherstone2018}. We use the standard spherical coordinates---$r$, $\theta$, and $\phi$---the local radius, colatitude, and azimuth, respectively---and the corresponding unit vectors $\hat{\bm{e}}_r$, $\hat{\bm{e}}_\theta$, and $\hat{\bm{e}}_\phi$. 

In addition to the anelastic approximation, which effectively filters out sound waves (e.g., \citealt{Gough1969}; \citealt{Gilman1981}; \citealt{Jones2011}), the thermodynamic variables are linearized about a temporally steady and spherically symmetric reference state, which is adiabatically stratified in the CZ ($d\overline{S}/dr=0$) and stably stratified in the RZ ($d\overline{S}/dr=0.014\ \rm{erg\ g^{-1}\ K^{-1} cm^{-1}}$, similar to the solar value; see \citealt{Brun2011}). The entropy gradient is smoothly matched between these constant values in the CZ and RZ over a radial thickness $0.05\rsun$. Hydrostatic balance then specifies the reference state uniquely. The pressure, density, temperature, and entropy are denoted by $P$, $\rho$, $T$, and $S$, respectively; overbars indicate the reference state and no overbars indicates the small perturbations about the reference state.

With these assumptions, the MHD equations are given by (e.g., \citealt{Matilsky2020a}; \citealt{Bice2020})
\begin{align}
\nabla\cdot(\overline{\rho}\bm{v}) = \nabla\cdot\bm{B} &=  0,
\label{eq:div}
\end{align}
\begin{align}
\overline{\rho}\Bigg{[}\frac{\partial\bm{v}}{\partial t} + (\bm{v}\cdot\nabla)\bm{v}\Bigg{]} = &-\overline{\rho}\nabla \Bigg{(}\frac{P}{\overline{\rho}}\Bigg{)} +\bigg{(}\frac{S}{\cp}\bigg{)}\overline{\rho}g\e_r + \nabla\cdot \bm{D}\nonumber\\
&-2\overline{\rho}\bm{\Omega}_0\times\bm{v} + \frac{1}{4\pi}(\curl\bm{B})\times\bm{B},
\label{eq:mom}
\end{align}
\begin{align}
\overline{\rho}\overline{T}\Bigg{[}\frac{\partial S}{\partial t} + \bm{v}\cdot\nabla S\Bigg{]} =\ &\nabla\cdot\big{[}\kappa\overline{\rho}\overline{T}\nabla S\big{]} + Q + \frac{\eta}{4\pi}|\curl\bm{B}|^2 \nonumber \\
&+ 2\overline{\rho}\nu\bm{D}\bm{:}\nabla\bm{v},
\label{eq:en}
\end{align}
\begin{align}
\pderiv{\bm{B}}{t}  =\ &\curl [\bm{v}\times\bm{B} - \eta\curl\bm{B}],
\label{eq:ind}
\end{align}
and
\begin{align}
\frac{\rho}{\overline{\rho}} = \frac{P}{\overline{P}} - \frac{T}{\overline{T}} = \frac{P}{\gamma\overline{P}}- \frac{S}{\cp}\ \text{with}\ \gamma=\frac{5}{3} ,\label{eq:eos}
\end{align}
Here, $\bm{v}$ is the fluid velocity, $\bm{B}$ the magnetic field, $\cp$ the constant-pressure specific heat, and $g=g(r)=GM_\odot/r^2$ the radially-varying gravitational acceleration.  We define $\Omega_\odot \equiv 2.87\times10^{-6}\ \rm{rad}\ \rm{s}^{-1}$ (the sidereal Carrington rotation rate), and our shell rotates at three times this rate to ensure a solar-like differential rotation and avoid the ``convective conundrum'' \citep{OMara2016}. We choose the diffusivities $\nu(r)$, $\kappa(r)$, and $\eta(r)$ to vary with radius like $\rhobar(r)^{-1/2}$. These diffusivities should be regarded as turbulent ``eddy'' diffusivities, which, for simplicity, we choose to appear in the MHD equations like molecular diffusivities, but with enhanced values. At the top of the domain, $\nu=\kappa=\sn{5}{12}\ \rm{cm^2\ s^{-1}}$ and $\eta=\sn{6.25}{11}\ \rm{cm^2\ s^{-1}}$, yielding Prandtl number values of ${\rm{Pr}}\equiv\nu/\kappa =1$ and $\prm\equiv\nu/\eta=8$ throughout. We use the internal heat source $Q=Q(r)$ to represent heating due to radiation, injecting a solar luminosity into the CZ. The heating is distributed, with highest values in the lower $\sim$1/3 of the CZ (see \citealt{Featherstone2016a}). At both boundaries, the velocity field is stress-free and impenetrable, the magnetic field is matched onto a potential field, and the radial entropy gradient is fixed, so that no conductive flux passes through the lower boundary, and the solar luminosity exits through the top boundary via conduction (see \citealt{Matilsky2020b}). We consider both an HD case, initialized from a random thermal field, and an MHD case, which is initialized by adding a small random magnetic ``seed'' field (amplitude $\sim$1 G) to the equilibrated HD case. Both cases are run until the kinetic (or magnetic) energy reaches a statistically steady state. In the MHD case, the magnetism is amplified by a convective dynamo to strengths of $\sim$10 kG. Overall, the dynamo is $\alpha\Omega$-like, similar to previous global simulations (e.g., \citealt{Guerrero2016a, Matilsky2020c}).

\section{Convection and Dynamo with Magnetically Confined Tachocline}\label{sec:tacho}
\begin{figure}
	\centering
	\includegraphics[width=0.85\linewidth]{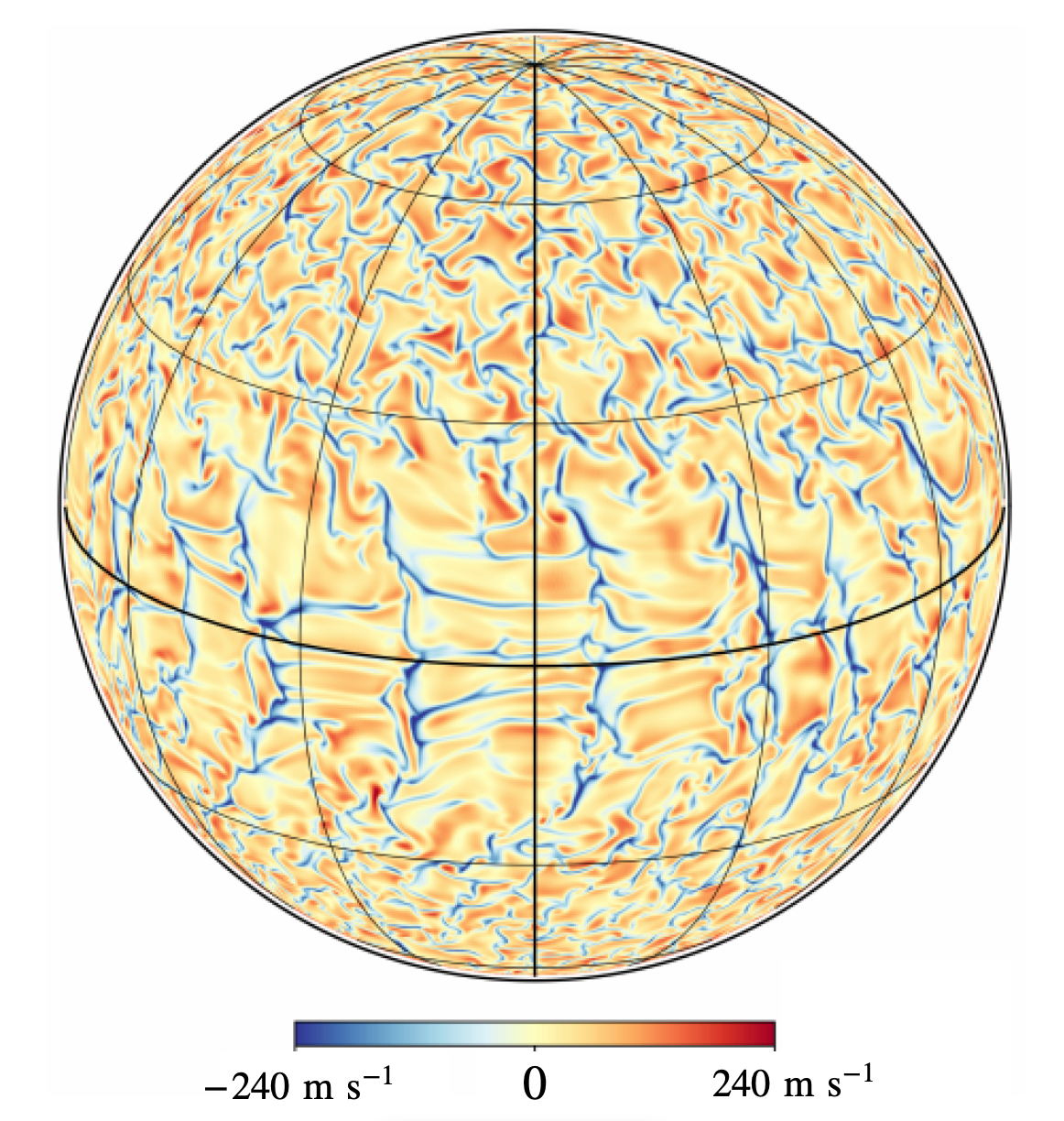}
	\caption{Orthographic projection of the typical flow field (the radial velocity $v_r$) achieved in a turbulent convection simulation with Rayleigh, shown on a spherical surface just below the outer boundary. Red tones indicate upflows ($v_r>0$) and blue tones indicate downflows ($v_r<0$). Parallels and meridians, separated by $30^\circ$, are marked by black lines, with the equator and central longitude in bold.}
	\label{fig:ortho}
\end{figure}

Spherical-shell convection simulations are highly nonlinear and turbulent, yielding intricate and time-dependent flow structures. In Figure \ref{fig:ortho}, we show a typical example of turbulent flow achieved in a simulation similar to the HD case explored here. Fairly orderly columnar structures, or ``Busse columns'' \citep{Busse2002}, are present at low latitudes. These columns are tilted in cylindrical cross-section and thus drive the solar-like differential rotation. At higher latitudes, the flow is more isotropic, consisting of small-scale vortices. Note that due to the density stratification, the downflows  ($v_r<0$) are significantly faster and narrower than the upflows ($v_r>0$). All flow structures evolve quite rapidly, maintaining their coherence for no longer than the dynamical timescale, $\sim$1 $\prot\equiv2\pi/\Omega_0\approx 8.4$ days. 

\begin{figure}
	\centering
	\includegraphics[width=0.85\linewidth]{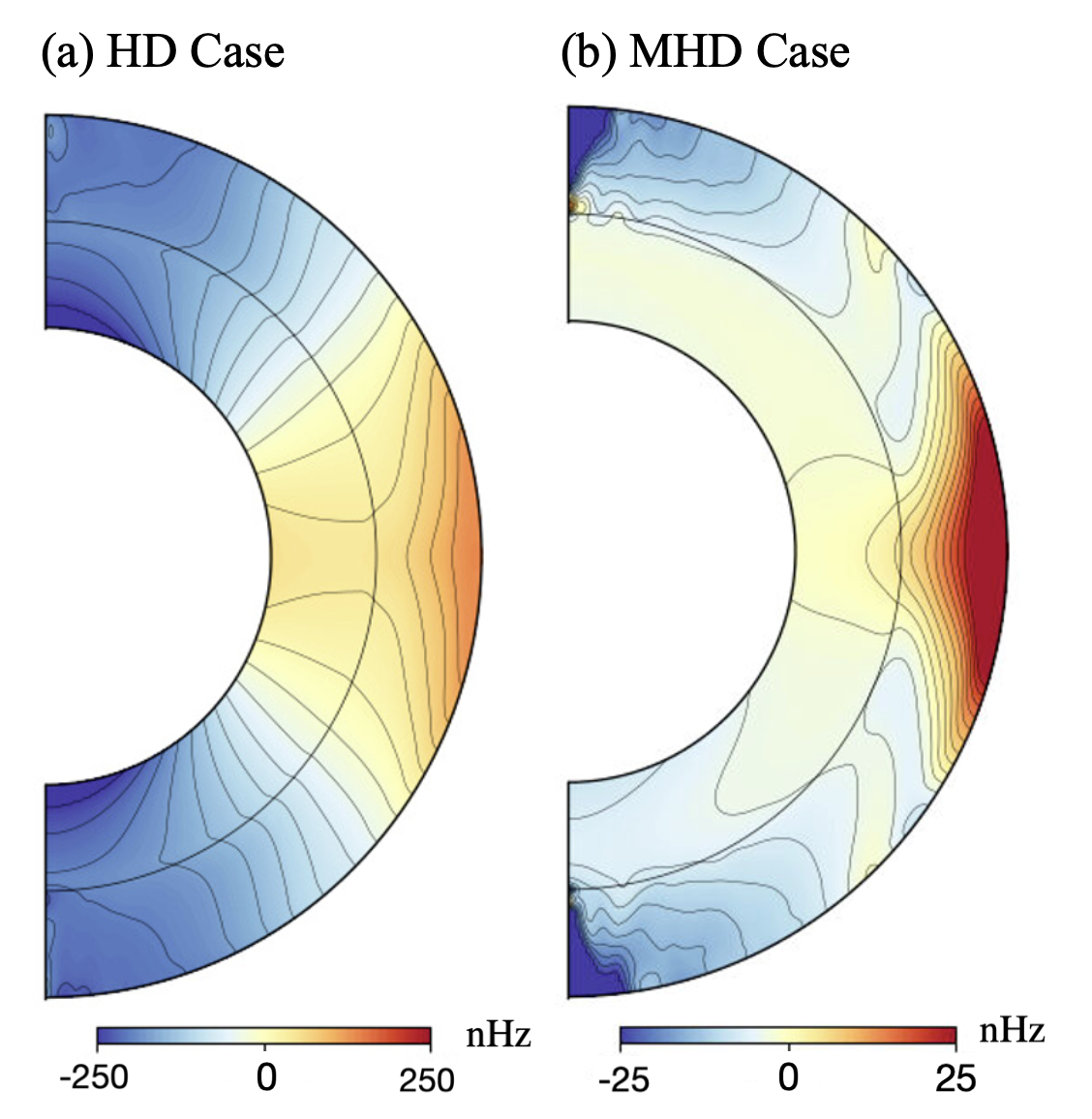}
	\caption{Rotation profiles the HD and MHD cases. The temporally and longitudinally averaged rotation rate as a function of $r$ and $\theta$ is shown in the full meridional plane. The middle circular line denotes $r=\rbcz=0.72\rsun$, the base of the CZ and roughly the top of the overshoot layer. In the HD case, a solar-like differential rotation has viscously imprinted from the CZ onto the entire RZ, but in the MHD case, the RZ is forced to rotate as a solid body, yielding a tachocline-like shear layer.}
	\label{fig:dr}
\end{figure}

We define the rotation rate in the non-rotating frame as $\Omega(r,\theta)\equiv \Omega_0 + \av{v_\phi}/r\sin{\theta}$, where the angular brackets denote a combined temporal and longitudinal average. In Figure \ref{fig:dr}, we show the rotation profiles achieved in the HD and MHD cases. The HD case supports a strong differential rotation, with equator rotating $\sim$40\% faster than the polar regions, a similar contrast to that of the Sun. The RZ rotates differentially like the CZ, which has viscously imprinted its differential rotation downward. In the MHD case, the overall differential rotation is substantially reduced (equator rotating only $\sim$5\% faster than the polar regions in the CZ). But most striking is the presence of clear solid-body rotation in the RZ, yielding a tachocline-like shear layer. 

\begin{figure}
	\centering
	\includegraphics[width=0.85\linewidth]{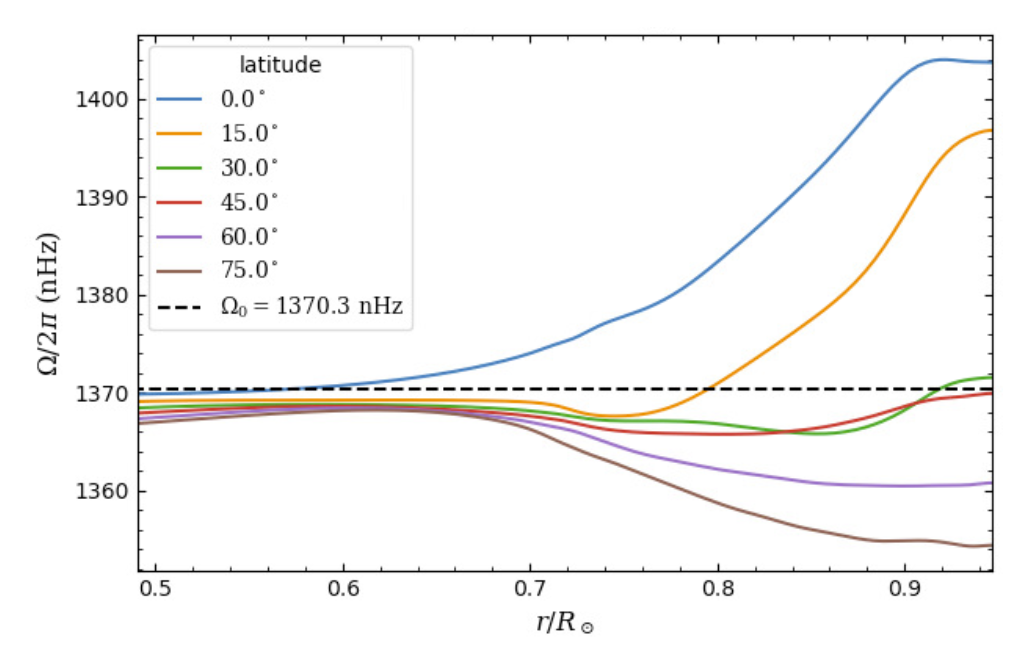}
	\caption{Rotation rate in the MHD case, shown at various latitudes as a function of radius, similar to Figure \ref{fig:solar}. The frame rate ($\Omega_0=3\Omega_\odot$) is shown as the black dashed line. The RZ rotates essentially like a solid body, while a small, but finite, differential rotation persists in the CZ, thus yielding a tachocline-like shear layer. }
	\label{fig:drr}
\end{figure}

We quantify this shear layer explicitly in Figure \ref{fig:drr}, which shows the MHD case's rotation profile along radial lines at various latitudes, similar to Figure \ref{fig:solar}. The greatest latitudinal contrast is achieved near the outer boundary. With increasing depth, the level of contrast decreases, until in the RZ, the curves have all tightened to the same value, again revealing solid-body rotation. We note that our tachocline is not thin, but takes up most of the CZ if we compare it to the solar one. In the Sun, the rotation rate is largely independent of radius (i.e., the rotation contours are tilted---see \citealt{Matilsky2020b}). This yields similar latitudinal contrasts at most depths above the tachocline, and stands in contrast the radially-dependent $\Omega(r,\theta)$ achieved in our MHD case. 

We explicitly define the rotation contrast at radius $r$ through the standard deviation of $\Omega(r,\theta)$ on the spherical surface of radius $r$. This deviation is 
\begin{align*}
\sigma_\Omega(r)\equiv \{\overline{[\Omega(r,\theta)-\overline{\Omega}(r)]^2}\}^{1/2},
\end{align*}
where the overbar indicates a spherical (latitudinal) average, ignoring latitudes above $60^\circ$. We define $\Delta\Omega(r)\equiv3\sigma_\Omega(r)$ to roughly match the difference in rotation rate at given radius between high latitudes and the equator. We thus define the \textit{solid-body ratio} of the RZ through $\domrz/\domcz$, where the subscripts indicate volume averages of $\Delta\Omega(r)$ over the RZ or CZ. For the CZ, we average only over the outer part ($r>0.9\rsun$) to capture the full magnitude of the differential rotation contrast achieved. 

In the MHD case, the solid-body ratio is $\sim$0.1. Given the limited spatial resolution of the helioseismic inversion kernels (e.g., \citealt{Howe2009}) the true solid-body ratio for the Sun is uncertain, but the solar RZ (to the depth helioseismology can probe, i.e. down to $r\sim0.5\rsun$) is ``at least as solid-body'' as $\domrz/\domcz\lesssim 0.07$. The rotation profile in the MHD case's RZ is thus on par with helioseismic constraints, and we now investigate the source of the solid-body rotation. 

\section{Ferraro's Law for Non-Axisymmetric Magnetic Fields}
Ferraro's law of isorotation \citep{Ferraro1937} is essentially the statement: ``Differential rotation in stellar RZs is only possible if the isorotation contours coincide with the poloidal magnetic field lines.'' In the notation of our simulations, this translates to the mathematical statement ``$\av{\bpol}\cdot\nabla\Omega=0$ in equilibrium,'' where $\bpol\equiv B_r\e_r + B_\theta\e_\theta$ is the poloidal magnetic field. Physically, Ferraro's law arises from the fact that a poloidal field ($\bpol$) in the presence of shear ($\nabla\Omega$)  will tend to produce a toroidal field according to the induction equation: $B_\phi\sim \bpol\cdot\nabla\Omega$. The associated magnetic tension force in the azimuthal direction will then be $\sim(\bm{B}\cdot\nabla\bm{B})_\phi\sim\bpol\cdot\nabla B_\phi\sim |\bpol|^2\nabla_{\rm{pol}}^2\Omega$, where $\nabla_{\rm{pol}}\equiv(\bpol/|\bpol|)\cdot\nabla$ is the derivative in the direction of $\bpol$. This force acts like a ``magnetic viscosity,'' tending to smooth out the rotation rate along $\bpol$-lines. Hence, a better statement of Ferraro's law might be: ``Poloidal magnetic field lines resist being bent by shearing motions.''

Both the magnetic confinement scenarios for the solar tachocline are primarily restatements of Ferraro's law. In the slow scenario \citep{Gough1998}, the relevant $\bpol$ is a primordial field confined to the RZ, and the magnetic-tension viscosity becomes significant on the long, Eddington-Sweet timescale. In the fast magnetic scenario \citep{ForgcsDajka2001}, the $\bpol$ is assumed to be the cyclic, convectively-generated dynamo field penetrating to a skin depth $\delta\sim \sqrt{\eta/\omega}$ below the base of the CZ. Here, $\omega\equiv2\pi/(22\ \text{years})$ and $\eta$ is a turbulently enhanced magnetic diffusion. In the fast scenario, the magnetic-tension viscosity becomes significant on the timescale of the cyclic solar dynamo, i.e., $\sim$22 years. 

If $\bpol$ is allowed to be non-axisymmetric, Ferraro's law becomes rather more restrictive (e.g., \citealt{Mestel1987}). The condition $\bpol\cdot\nabla\Omega$ must now be satisfied locally, and since $\bpol$'s direction changes with longitude, the only way to achieve this is through $\Omega=\text{constant}$, i.e., solid-body rotation. As shown below, our MHD case is an example of fast magnetic tachocline confinement (in a global, 3D geometry and including convection) with a non-axisymmetric $\bpol$. 

\begin{figure*}
	\centering
	\includegraphics[width=0.85\linewidth]{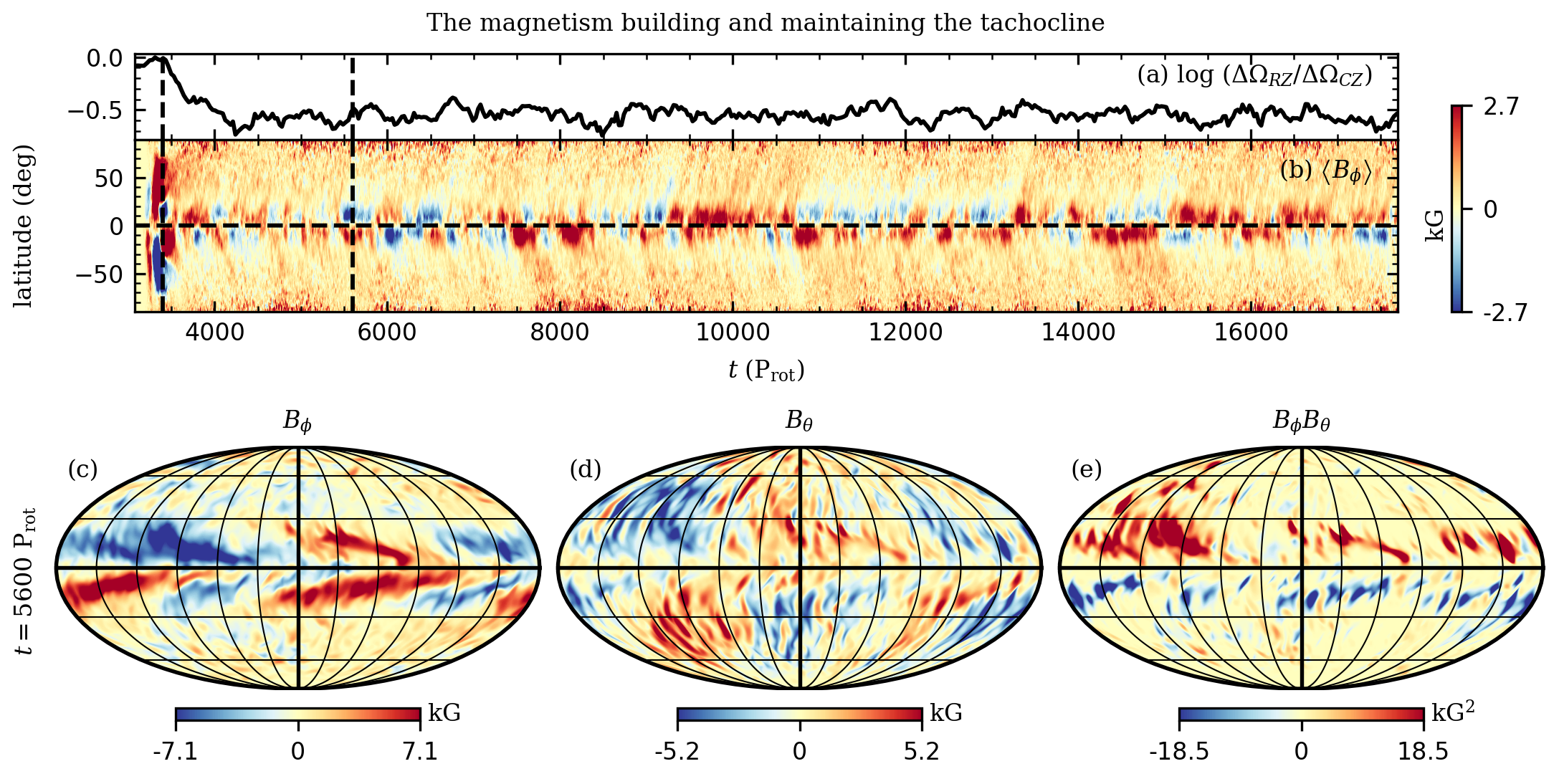}
	\caption{(a) Temporal evolution of the solid-body ratio in a simulation identical to the MHD case, but with $\prm=4$. Initially (first vertical dashed line) the RZ rotates differentially like the CZ (as in the HD case), but eventually the RZ is forced to rotate like a solid body. (b) Time-latitude diagram of the longitudinally averaged azimuthal magnetic field $\av{B_\phi}$. (c--e) Mollweide projections of $B_\phi$, $B_\theta$, and the product $B_\phi B_\theta$ (respectively), on a spherical surface just below the overshoot layer ($r/\rsun=0.69$) at $t=5600\ \prot$, the instance marked by the second vertical dashed line in panels (a, b).}
	\label{fig:moll}
\end{figure*}

In Figures \ref{fig:moll}(a, b), we show the temporal evolution of a magnetic simulation similar to the MHD case (the only difference here being $\prm=4$ instead of $\prm=8$). Figure \ref{fig:moll}(a) shows the solid-body ratio (where this time the temporal average is not included in our computation of $\Omega$) evolving from its initial value near unity (inherited from the HD case, in which the RZ and CZ have similar differential rotation contrasts) to a much lower value, in which the RZ rotates like a solid-body. Note that due to the lack of temporal averaging in $\Omega$, the instantaneous solid-body ratio hovers around the value $\sim$0.25, slightly higher than the $\sim$0.1 value reported earlier for the temporally averaged $\Omega$. Figure \ref{fig:moll}(b) shows the time-latitude diagram of $\av{B_\phi}$. Just before the solid-body ratio plummets (first vertical dashed line), $\av{B_\phi}$ is strong and largely axisymmetric. We have verified that the magnetic tension associated with the axisymmetric poloidal field lines directly leads to the elimination of differential rotation in the RZ. There are some ostensible cycles for the rest of the simulation, but they are quite chaotic and difficult to interpret, since the primary magnetic field is non-axisymmetric, and cycles in $\av{B_\phi}$ do not necessarily imply polarity reversals in the usual sense. 

Figures \ref{fig:moll}(c--d) show the non-axisymmetric magnetic field structure responsible for maintaining the tachocline in a steady state. We have chosen to view the horizontal field ($B_\phi$, $B_\theta$, and the product $B_\phi B_\theta$) on a spherical surface just below the overshoot layer ($r/R_\odot=0.69$) at a random time in the simulation (second dashed line in Figures \ref{fig:moll}(a, b)). Both field components follow the same spatial distribution overall: bisymmetry ($m=1$) in the North and quadrisymmetry ($m=2$) in the South. The fields are tightly correlated so that their product $B_\phi B_\theta$ has a preferred sign structure of negative in the North and positive in the South. Over the course of the simulation, these structures vary a bit, mostly by switching the primary magnetism---in each hemisphere separately---between the $m=1$ and $m=2$ modes. However, the correlation $B_\phi B_\theta$ always remains very strong, with an overall profile similar to the one shown in Figure \ref{fig:moll}(e). 

This correlation represents the two main ingredients of Ferraro's law. First, $B_\phi$ is generated from $B_\theta$ inductively by the mean shear: for a solar-like differential rotation (even if quite small), the mean shear term $\bpol\cdot\nabla\Omega$ will always have the same sign as $B_\theta$ in the North and the opposite sign in the South. Second, this sign structure of $B_\phi B_\theta$ represents a \textit{poleward} angular momentum transport by magnetic tension forces (note that the horizontal angular momentum flux is proportional to \textit{minus} $B_\theta B_\phi$---see \citealt{Miesch2011}). This magnetic transport thus attempts to eliminate the latitudinal shear. We have verified that the small amount of residual shear in the RZ depicted in Figures \ref{fig:dr} and \ref{fig:drr} is due to the balance of this magnetic tension force and the viscosity. This balance persists over $\sim$10 magnetic diffusion times across the RZ, thus indicating a steady-state shear layer. Overall, our MHD case seems to be a realization of the fast magnetic confinement scenario (notably confinement against viscous spreading and not radiative spreading), but with greatly different field geometry (non-axisymmetry) and no clear cycling behavior. 

\section{The Solar Tachocline}
We have presented the first 3D, global, spherical-shell simulation of a CZ--RZ system to achieve one of the three main tachocline confinement scenarios (fast magnetic). In \citet{Barnabe2017}, the fast magnetic scenario was shown to be effective at preventing radiative (as well as viscous) spreading of the tachocline in a 1D model without convection, even in the low turbulence limit $\eta\sim10^7\rm{cm^2\ s^{-1}}$. Because of numerical constraints, we are not able to run our model for the simulation's Eddington-Sweet time (in the MHD case, $\sim$$\sn{2}{7}$ rotations) or at values of the turbulent magnetic diffusivity lower than $\eta\sim10^{11}\ \rm{cm^2\ s^{-1}}$. Nevertheless, our MHD case is a proof-of-concept that the fast magnetic scenario can function in a global geometry, with the RZ coupled to convective overshoot and the dynamo self-consistently generated by the convection. 

Although we cannot say with certainty how our MHD case may react to the presence of primordial magnetic fields, we note a striking feature of the application of Ferraro's law achieved here. Our convective dynamo generates a global magnetic field, which couples the CZ to the RZ. Nonetheless, Ferraro's law does \textit{not} imprint the differential rotation of the CZ onto the RZ, and instead forces the RZ to rotate like a solid-body. This stands in contrast to the models of \citet{Strugarek2011a, Strugarek2011b}, who ran global Anelastic Spherical Harmonic (ASH) simulations of a coupled CZ--RZ system with a primordial magnetic field initially confined to the RZ. It was found that through three-dimensional convective instabilities, the magnetic field ``leaked'' into the CZ, and the geometry was such that Ferraro's law imprinted the differential rotation of the CZ onto the RZ along the poloidal magnetic field lines. %In our MHD case, of course, some of the same imprinting occurs, but in reverse. Solid-body rotation in the RZ couples via the global poloidal field to greatly reduce the rotation contrast in the CZ. Further work must there assess how Ferraro's law is prevented from being a primary player in the CZ.

The main difference between those prior models and our MHD case is that in general $|B_\theta|\sim5|B_r|$ in the overshoot layer, thus implying a fairly horizontal poloidal magnetic field. Ferraro's law thus eliminates \textit{latitudinal} shear much more efficiently than \textit{radial} shear. This anisotropy of ``magnetic-tension viscosity'' evidently can produce a tachocline-like shear layer dominated by radial shear, as shown in Figure \ref{fig:drr}. It is worth noting that the solar tachocline (although significantly different from our MHD case in that it is very thin and has much stronger rotation contrast in the CZ) is also a region in which the radial shear is substantially greater than the latitudinal shear. We have not done an exhaustive analysis of the $\alpha$-effect responsible for this horizontal poloidal magnetic field, and so cannot comment whether it might be a general feature of magnetic convective overshoot regions (in other dynamo simulations, or in the Sun). However, we note that in both magnetic confinement scenarios, the radial field component was ignored, largely for convenience in the analytical calculations. It is thus rather intriguing that our MHD case generates a mostly horizontal poloidal magnetic field in the overshoot layer self-consistently.

\section*{Acknowledgments}
{The authors thank Sacha Brun, Antoine Strugarek, Connor Bice, and Bradley Hindman, for helpful conversations on tachocline confinement and Ferraro's law. Loren I. Matilsky was partly supported during this work by the NASA FINESST award 80NSSC19K1428. This research was primarily supported by NASA Heliophysics through grants 80NSSC18K1127, NNX17AG22G, and NNX13AG18G. Computational resources were provided by the NASA High-End Computing (HEC) Program through the NASA Advanced Supercomputing (NAS) Division at Ames Research Center. Rayleigh has been developed by Nicholas Featherstone with support by the National Science Foundation through the Computational Infrastructure for Geodynamics (CIG) under NSF grants NSF-0949446 and NSF-1550901.}

\bibliographystyle{cs20proc}
%\bibliography{zz_my_library.bib}

\bibliography{my_library2.bib}
%\bibliography{/Users/loren/Desktop/00_Research_Papers/zz_my_library}

\end{document}